\documentclass[12pt,a4paper]{article}

\usepackage[american]{babel}
\usepackage[a4paper,top=2cm,bottom=2cm,left=2cm,right=2cm,marginparwidth=1.75cm]{geometry}
\usepackage[style=apa, backend=biber]{biblatex}
\usepackage{amsmath, graphicx, float, booktabs, caption, authblk, setspace, placeins, titlesec, lineno, listings}

\usepackage{colortbl}
\usepackage[table]{xcolor}
\usepackage{multirow, multicol}
\usepackage{booktabs}
\usepackage{geometry}

\usepackage{xcolor}
\definecolor{MyBlue}{HTML}{6699ff}

\usepackage{hyperref}
\hypersetup{
  colorlinks   = false,  
  linkbordercolor = MyBlue,        
  citebordercolor = MyBlue,        
  urlbordercolor  = MyBlue,        
  pdfborderstyle={/S/S/W 1}      
}

\DefineBibliographyStrings{english}{techreport = {}}

\usepackage{arydshln}
\usepackage{float}
\usepackage{siunitx}
\usepackage{arydshln}
\usepackage{siunitx}
\usepackage{array}
\usepackage[none]{hyphenat}
\usepackage{caption}
\DeclareCaptionFont{captionsmall}{\small}
\title{AI labeling reduces the perceived accuracy of online content but has limited broader effects\thanks{We thank Verian UK Limited for providing the sample used in this study. We are grateful for the valuable feedback received at the APSA Annual Meeting (2024, 2025), particularly from Andrew Guess (discussant) and Leticia Bode (chair). We also appreciate the helpful suggestions from participants in seminars at the LSE Department of Methodology and the LSE Data Science Institute.

Author contributions: C.W. designed the research, conducted the data analysis, wrote the first draft, and revised the manuscript. P.S. designed the research, advised on analysis, and revised the manuscript. D.d.K. designed the research, advised on analysis, and revised the manuscript. All authors approved the final manuscript. C.W. is a PhD candidate in the Department of Methodology at the London School of Economics and Political Science (LSE). P.S. is Professor of Quantitative Social Science at LSE. D.d.K. is an Assistant Professor of Quantitative Research Methodology at LSE.
}}

\author[1,2]{Chuyao Wang\thanks{Corresponding author, email: c.wang85@lse.ac.uk}}
\author[1]{Patrick Sturgis}
\author[1,2]{Daniel de Kadt}
\affil[1]{\small Department of Methodology, London School of Economics and Political Science}
\affil[2]{\small Data Science Institute, London School of Economics and Political Science}

\date{} 

\begin{document}
\maketitle

\begin{abstract}
Explicit labeling of online content produced by artificial intelligence (AI) is a widely discussed policy for ensuring transparency and promoting public confidence. Yet little is known about the scope of AI labeling effects on public assessments of labeled content. We contribute new evidence on this question from a survey experiment using a high-quality nationally representative probability sample (\emph{n} = 3,861). First, we demonstrate that explicit AI labeling of a news article about a proposed public policy reduces its perceived accuracy. Second, we test whether there are spillover effects in terms of policy interest, policy support, and general concerns about online misinformation. We find that AI labeling reduces interest in the policy, but neither influences support for the policy nor triggers general concerns about online misinformation. We further find that increasing the salience of AI use reduces the negative impact of AI labeling on perceived accuracy, while one-sided versus two-sided framing of the policy has no moderating effect. Overall, our findings suggest that the effects of algorithm aversion induced by AI labeling of online content are limited in scope.
\end{abstract}

\begin{center}
\small{\textbf{Keywords}: AI Labeling; Public Opinion; Survey Experiment; Generative AI; Algorithm Aversion}
\end{center}


\newpage

\section{Introduction}
\label{sec:intro}
The rapid development of generative artificial intelligence (AI) has made AI-generated content (AIGC) largely indistinguishable from human-written text \parencite{jakesch_hancock_naaman_2023,kobis_mossink_2021}. These developments have raised concerns among policymakers about the potential negative impacts of AIGC on public trust and the proliferation of misinformation online \parencite{garimella_chauchard_2024,kozyreva_lorenzspreen_herzog_ecker_lewandowsky_hertwig_wineburg_2024,spitale_billerandorno_germani_2023}. AI labeling, which requires public disclosure of AI involvement in generating online content, is an intuitively appealing policy solution to address this problem \parencite{CAC2025, european_parliament_2024,us_congress_2023}. However, some experimental studies have found that labeling texts as AI-generated, rather than human-written, reduces their perceived accuracy, even when the content is identical \parencite{altay_gilardi_2024,chen_wang_hao_2025,palmer_spirling_2023}. AI labeling may thus decrease public trust in factually accurate AIGC or increase trust in factually inaccurate online content that is unlabeled, and so is assumed to be human-written \parencite{henestrosa_greving_kimmerle_2023,palmer_spirling_2023}. These would be unintended and undesirable outcomes.

Why might people perceive text to be less accurate simply because it is labeled as having been produced by an AI? One compelling explanation for this phenomenon is that the AI labeling effect is an example of algorithm aversion, where people have been found to trust algorithmic outputs less than human ones in decision-making contexts \parencite{dietvorst_simmons_massey_2015,dietvorst_simmons_massey_2018}. Algorithm aversion tends to occur when prior algorithmic errors are made salient and when algorithms are used to decide on subjective or evaluative tasks \parencite{mahmud_islam_ahmed_smolander_2022}. By highlighting AI involvement in the task of content creation, AI labeling may therefore cause public concern about the quality and accuracy of AIGC, through algorithm aversion, even when the content is factually accurate.

Empirically, experiments using non-probability samples have shown that AI labeling can shape initial perceptions of the labeled content, such as perceived accuracy \parencite{altay_gilardi_2024,chen_wang_hao_2025}, willingness to share content \parencite{altay_gilardi_2024}, and the favorability of arguments \parencite{palmer_spirling_2023}. A key question now is whether these effects extend beyond such accuracy and credibility evaluations. In a similar context, research has found that labeling text as misinformation reduces trust not only in the labeled content but also in unrelated information, suggesting that labeling can have broader contamination effects \parencite{hameleers_2023, vandermeer_hameleers_ohme_2023}. This raises the possibility that AI labeling may also generate limited spillover effects, whereby changes in content-level credibility evaluations extend to other judgments, including reduced interest in the labeled information, lower support for the arguments presented, or heightened general concern about misinformation. We investigate these possibilities here, using a factorial survey experiment on a large, nationally representative probability sample of UK adults.

We offer two novel contributions to the AI labeling literature. First, we replicate the effect of AI labeling on perceived information accuracy using a high-quality sample. We present respondents with an article about a major public policy proposal and manipulate the presence/absence of AI labeling. Second, we assess potential spillover effects on interest in and support for the policy presented, as well as on general concern about misinformation. This shows that AI labeling reduces interest in the described policy but has no statistically significant effect on policy support or concerns about online misinformation.

Beyond this, we offer two further insights. First, given the limited public awareness of generative AI during the study period, we assess whether first increasing the salience of AI use in text generation moderates the effect of AI labeling on respondent assessments. Our results show that enhancing the salience of generative AI use reduces the negative effect of AI labeling on perceived accuracy but has no moderating influence on the other three outcomes. Second, existing research has focused on the impact of AI labeling on ostensibly “neutral” information. Yet much online AI-generated content promotes a particular stance or policy position. We, therefore, test if the effect of AI labeling is moderated by whether the framing is one-sided (advocacy) or two-sided (balanced), finding no statistically significant effect.


\section{Relevant Literature and Hypotheses}
\label{sec:hyp}

AI labeling can function as a disclosure cue that meaningfully shapes human judgment. In particular, labels such as “AI-generated” versus “human-generated” can shift evaluations even when the underlying content is held constant \parencite{zhu2025human}. Related research highlights that disclosure wording and presentation structure shape how AI labels are understood and contextualized by users, which makes it important to clearly specify the labeling format employed in any given study \parencite{wittenberg2024labeling,gamage2025labeling}. In this study, we focus on a single format of AI labeling (see Appendix \ref{sec:A_labeling}) that aligns with AI labeling regulations \parencite{CAC2025, european_parliament_2024,us_congress_2023}. Below, we organize the literature review around three related strands: AI labeling's effect on perceived accuracy, potential spillover effects beyond perceived accuracy, and contextual moderators that may condition AI labeling effects.

\subsection{AI Labeling's Effect on Perceived Accuracy}

Empirically, randomized experiments have shown that AI labeling reduces the perceived accuracy of online content \parencite{altay_gilardi_2024,longoni_fradkin_cian_pennycook_2022} and assessments of content quality \parencite{chen_wang_hao_2025,palmer_spirling_2023}. Yet, other experimental studies have found no statistically significant impact on the labeled content’s perceived credibility \parencite{henestrosa_greving_kimmerle_2023} or quality \parencite{zhang_gosline_2023}. These mixed findings may be partly due to the reliance on 'opt-in' non-probability samples in existing research. Although random assignment of treatments provides unbiased estimates of causal effects within each sample, the potential for heterogeneous treatment effects across non-probability samples may limit the generality and comparability of findings across studies \parencite{cornesse_et_al_2020}. To address this limitation, we test the effect of AI labeling on perceived information accuracy using a high-quality probability sample of the UK adult population, labeling the text for a half of the sample as AI-generated, with the other half seeing unlabeled text. Our first hypothesis (H1) is:

\vspace{12pt}
\textit{\textbf{H1}: The perceived accuracy of the article will be lower when it is labeled as AI-generated than when it is not.}
\vspace{12pt}

\subsection{Beyond Perceived Accuracy: AI Labeling's Potential Spillover Effects}

While several studies have considered the effect of AI labeling of articles on perceived accuracy, little is known about whether AI labeling produces effects that extend to other assessment dimensions. Respondents in our study evaluate a news article about a proposed public policy (Universal Basic Income, UBI), allowing us to test policy-relevant spillover effects of AI labeling. If AI labeling leads individuals to perceive labeled information as less accurate, it is plausible that they will also find it less engaging and less persuasive \parencite{petty_cacioppo_1986,chang_lu_lin_2020}. In policy communication, these downstream responses are often reflected as variation in issue-level engagement and evaluative support for the proposal itself, capturing distinct but policy-relevant dimensions of how citizens respond to policy information \parencite{muradova2020climate, yang2023institutions}. Accordingly, we examine whether AI labeling affects expressed interest in and support for the described policy. This motivates our second hypothesis (H2):

\vspace{12pt}
\textit{\textbf{H2}: Interest in, and support for the policy proposal will be lower when the text is labeled as AI-generated than when it is not.}
\vspace{12pt}

Next, we consider whether AI labeling affects wider concerns about misinformation in the online environment. This may happen if the effect of AI labeling of text is to cue respondents to consider that this type of potentially inaccurate text is becoming widely prevalent online, similar to how misinformation labeling of specific text increases general distrust in the information environment \parencite{hameleers_2023,vandermeer_hameleers_ohme_2023}. In this sense, general misinformation concern provides a broader test of spillover, whereby evaluations of a specific labeled article generalize to beliefs about the credibility of the broader online information environment. Hypothesis three (H3) is then:

\vspace{12pt}
\textit{\textbf{H3}: General concern about misinformation in the online environment will be greater when the text is labeled as AI-generated than when it is not.}
\vspace{12pt}

\subsection{AI Labeling's Potential Moderators: Salience and Message Framing}

Beyond the labeling cue itself, the impact of AI labels may depend on whether AI involvement is cognitively salient to respondents. As of mid-2024, when this study was conducted, the salience of generative AI amongst the UK public was quite low: only 58\% had heard of ChatGPT, and just 7\% used it weekly \parencite{fletcher_nielsen_2024}. Another contemporaneous survey found just 40\% of adults reported high or moderate awareness of AI \parencite{arguedas_2024}. When respondents have low or no awareness of what generative AI is, it is not credible to assume that their assessments will be affected by the AI labeling treatment. We therefore manipulate the salience of AI by first presenting a random half of the sample with information about what generative AI is and how it is being used to generate online content, while the control group is not presented with any information to enhance salience. We expect the main effect of salience enhancement to produce effects in the same direction as AI labeling. Hypothesis four (H4) is thus:

\vspace{12pt}
\textit{\textbf{H4}: Perceived accuracy, policy interest, and policy support will be lower, while general misinformation concern will be higher when the salience of AIGC is enhanced than when it is not.}
\vspace{12pt}

The enhanced salience of AIGC may also augment the main effect of AI labeling, such that there is an interaction between salience and AI labeling. Specifically, salience enhancement may direct attention to AI labeling and activate pre-existing negative connotations of algorithmic involvement \parencite{sundar_2008}. This reasoning motivates the fifth hypothesis (H5):

\vspace{12pt}
\textit{\textbf{H5}: Perceived accuracy, policy interest, and policy support will decrease more, while general misinformation concern will increase more when AI labeling is combined with salience enhancement than when it is not.}
\vspace{12pt}

It is plausible that the effects of AI labeling will also be greater when the labeled text promotes a particular position (one-sided framing) than when it presents information in a balanced way (two-sided framing). In political communication research, message framing is known to shape perceptions of persuasive intent and credibility, motivating an assessment of whether AI labeling effects vary across framing contexts. One-sided messages are generally considered less trustworthy than two-sided messages \parencite{hendriks_janssen_jucks_2023}. This trust penalty may be compounded when the text is revealed to have been generated by AI. This intuition is supported by \textcite{henestrosa_greving_kimmerle_2023}, who found that members of the public rated AI-generated and human-written texts as similarly credible when neutrally presented. Yet, when texts with positive or negative framings were presented, their perceived credibility was lower, particularly for those attributed as being produced by AI. This leads us to expect an interaction between AI labeling and the one-sided (versus two-sided) framing of the proposed policy. Hypothesis six (H6) is then:

\vspace{12pt}
\textit{\textbf{H6}: Perceived accuracy will decrease more, while general misinformation concern will increase more when AI labeling is combined with a one-sided framing, compared to a two-sided framing.}
\vspace{12pt}

Finally, we test whether the combined effects of AI labeling, salience enhancement, and one-sided framing are multiplicative; that is, we expect that the four outcome assessments would be lowest for respondents exposed to all three treatments. AI labeling signals algorithmic control, whereas one-sided framing conveys persuasive intent. Salience enhancement may further intensify their impact by directing cognitive focus toward the constructed and strategic nature of AI-generated messages \parencite{sundar_2008}. When all three signals are present, salience enhancement may reinforce skepticism toward algorithmic authorship and strategic intent. This possibly activates a resistance schema and lowers credibility judgments, leading to our seventh hypothesis (H7):

\vspace{12pt}
\textit{\textbf{H7}: Perceived accuracy will be lowest, while general misinformation concern will be highest when AI labeling, salience enhancement, and a one-sided framing are combined.}
\vspace{12pt}


\section{Data and Methods}
\label{sec:data}
To test these hypotheses, we implemented a $2\times2\times2$ between-subject experimental design in which we manipulated three binary factors: (1) salience enhancement versus no salience enhancement, (2) AI labeling versus no AI labeling, and (3) one-sided versus two-sided framing. Participants were randomly assigned to one of eight equal-sized experimental conditions, as shown in Table \ref{table:design} below. Covariate balance checks indicated good balance across treatment groups, with all standardized mean differences well below 0.1.

\begin{table}[H]
\centering
\scriptsize{
\resizebox{\linewidth}{!}{
\renewcommand{\arraystretch}{2}
\begin{tabular}{c c c c c c c c c}
\toprule
\multicolumn{9}{c}{Random Assignment into 8 Groups} \\
& Group 1 & Group 2 & Group 3 & Group 4 & Group 5 & Group 6 & Group 7 & Group 8 \\
\midrule
Treatment 1 & \multicolumn{4}{c}{\cellcolor{gray!60}Salience Enhancement} & \multicolumn{4}{c}{\cellcolor{gray!25}No Enhancement} \\
Treatment 2 & \multicolumn{2}{c}{\cellcolor{gray!60}AI Labeling} & \multicolumn{2}{c}{\cellcolor{gray!25}No Labeling} & \multicolumn{2}{c}{\cellcolor{gray!60}AI Labeling} & \multicolumn{2}{c}{\cellcolor{gray!25}No Labeling} \\
Treatment 3 & \cellcolor{gray!60}One-sided & \cellcolor{gray!25}Two-sided & \cellcolor{gray!60}One-sided & \cellcolor{gray!25}Two-sided & \cellcolor{gray!60}One-sided & \cellcolor{gray!25}Two-sided & \cellcolor{gray!60}One-sided & \cellcolor{gray!25}Two-sided \\
\midrule
N & 448 & 477 & 530 & 458 & 489 & 484 & 486 & 489 \\
\bottomrule
\end{tabular}
\renewcommand{\arraystretch}{1.0} 
}
}
\captionsetup{justification=centering}
\caption{Experimental Design.}
\label{table:design}  
\captionsetup{justification=raggedright,singlelinecheck=false}
\caption*{\footnotesize{\textbf{Note}: Eight groups are formed by crossing three binary treatment conditions ($2 \times 2 \times 2$). By design, every respondent had an equal probability of being assigned to each group. Realized sample size in groups is shown.}}
\end{table}

The experiment proceeded as follows. Given limited public familiarity with generative AI at the time of data collection, participants were first assigned either to a salience enhancement condition or to a control condition (where nothing was shown). Those assigned to the former condition read a brief text explaining that generative AI can produce human-like writing and realistic images and that AI is increasingly used to produce news articles. They were then asked: “Over the past few months, how much have you heard or read about generative AI?”. In combination, this served as both a salience enhancement and as a baseline measure of AI familiarity at the time the study was conducted (see Appendix section \ref{sec:A_salience} for full wordings).

Next, all participants were asked to read a news article about a proposed public policy—Universal Basic Income (UBI). The article was formatted to resemble a BBC News webpage \parencite{mcnamee_2023}, complete with an image to match a typical digital news layout. Participants in the one-sided framing condition viewed an article titled “UBI: A Brighter Future for All in the UK”, which emphasized potential benefits such as better living standards (see \ref{sec:A_one-sided}). Participants in the two-sided framing condition were presented with an article titled “UBI: A New Horizon or a Step Too Far?”, which presented both the potential benefits such as supporting job creation and costs such as driving inflation (see \ref{sec:A_two-sided}).

For participants assigned to the AI labeling condition, the news article ended with a labeling element. This consisted of a sentence that is compliant with current and proposed regulations \parencite{CAC2025, european_parliament_2024,us_congress_2023} (“This report was generated by ChatGPT, an artificial intelligence”) and a simple circular “AI” icon (see Appendix \ref{sec:A_labeling}). The AI labeling condition was intended to indicate AI involvement in producing the report (text and the embedded image) as presented, while leaving the article text and layout otherwise identical across conditions. All participants then completed four questions on Likert scales on which they assessed: the perceived accuracy of the information, their level of interest in learning more about UBI, their likelihood of supporting UBI, and their extent of concern about misinformation in online news media (see \ref{sec:A_survey}).

The experiment was administered to a random probability sample of 3,861 British adults, drawn from the Verian UK \emph{Public Voice} online panel.\footnote{\textbf{Note}. Ethical approval for the experiment was obtained on 26 March 2024. The pre-analysis plan was preregistered on the Open Science Foundation on 15 May 2024 (\href{https://osf.io/5grkz/}{Link}). The survey was subsequently fielded online in mid-2024. Informed consent was provided with an online tick-box, and Verian anonymized all data to protect privacy.} \emph{Public Voice} uses address-based probability sampling to construct and maintain a panel of respondents who have agreed to take part in surveys periodically for small monetary incentives. The target population is individuals aged 16 or older, resident in Great Britain for at least five years, all of whom have a known, non-zero probability of selection from the Postcode Address File. Full details of the sample design and methodology can be found in the survey’s technical report \parencite{kantar_2022}.

To estimate treatment effects, we fit weighted least squares (WLS) regressions with robust standard errors using the \texttt{svyglm} package in \emph{R}. We apply weights provided by Verian that combine a design weight, a non-response weight, and a post-stratification weight. As a result, our regressions meaningfully characterize a population average treatment effect for the British public, but the choice to use weights does not alter our substantive or statistical conclusions \parencite{franco_malhotra_simonovits_zigerell_2017,miratrix_sekhon_theodoridis_campos_2018}. As detailed in Appendix \ref{sec:appendixB}, our regression analysis includes: (a) three specifications, each with only one factor corresponding to each of the three treatments (AI labeling, salience enhancement, or one-sided framing); (b) three specifications, each including two treatment factors and their corresponding two-way interaction; and (c) a full specification, including all three treatment factors and the two-way and three-way interactions (AI labeling × salience enhancement × one-sided framing). The main weighted least squares (WLS) estimates are reported in Appendix~\ref{sec:appendixC}. In addition, estimates from unweighted ordinal least squares (OLS) are included in Appendix \ref{sec:appendixD}, and the findings remain substantively unchanged. Finally, we fitted weighted ordinal logistic regression (OLR) models that preserve the ordinal structure of the outcomes (Appendix \ref{sec:appendixE}), yielding results that are consistent with the WLS estimates in both direction and statistical significance. 


\section{Results}
\label{sec:results}

Figure \ref{fig:awareness} below presents the baseline distribution of awareness of generative AI in our sample. Consistent with contemporaneous surveys in mid-2024 \parencite{arguedas_2024,fletcher_nielsen_2024}, awareness was quite low, with 53\% of respondents reporting having heard a small amount, not very much, or nothing at all about AI, or that they do not know or are not sure. This low public awareness is consistent with our motivation for testing whether increasing the salience of AIGC influences responses to subsequently presented text.

\begin{figure}[H]
  \centering
  \includegraphics[width=0.9\textwidth]{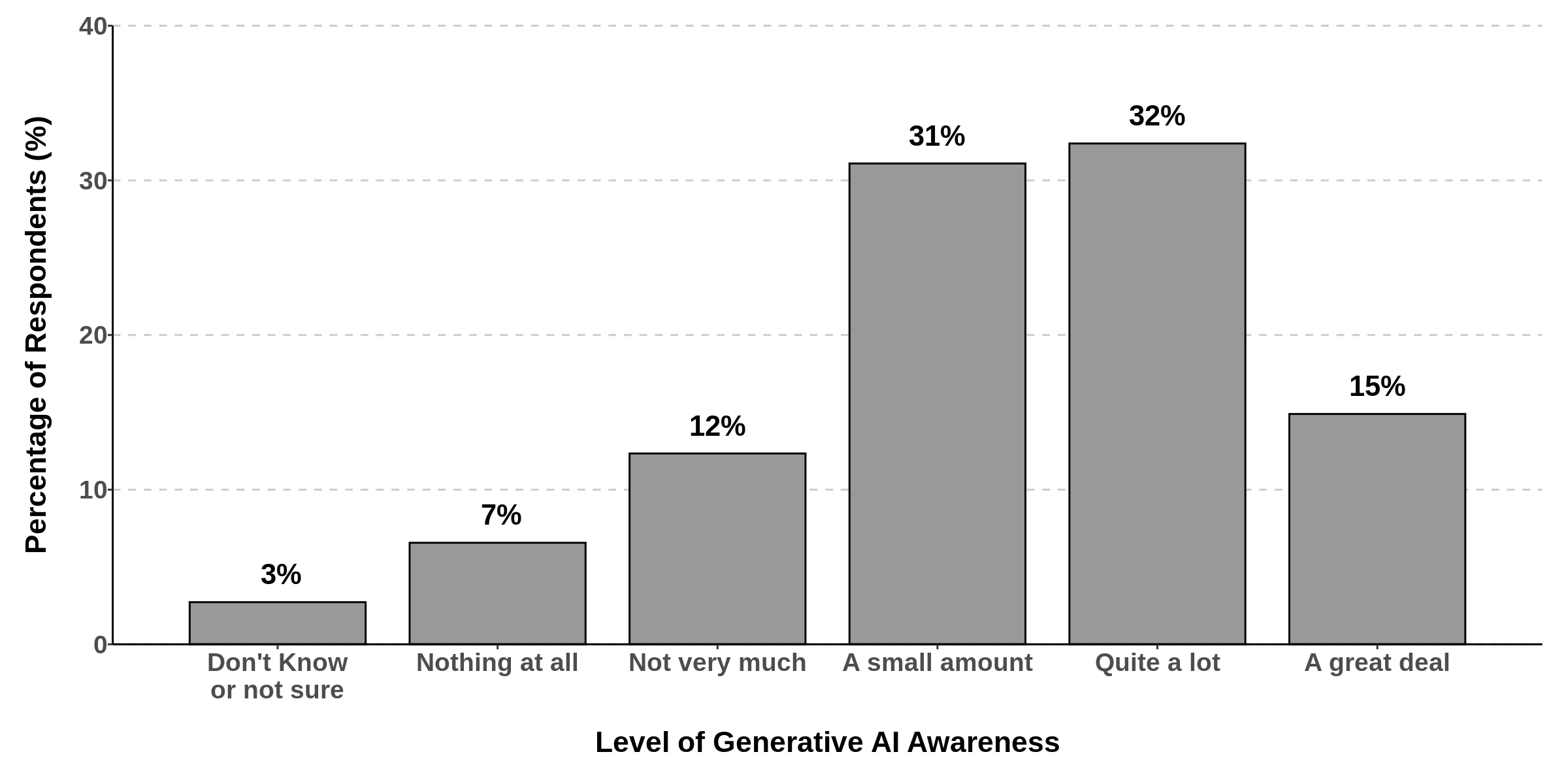}

  \captionsetup{justification=centering,singlelinecheck=true}
  \caption{Baseline Awareness of Generative AI in the Sample.\label{fig:awareness}}
  \captionsetup{justification=raggedright,singlelinecheck=false}
  \caption*{\footnotesize{\textbf{Note}}: Based on responses to the question “Over the past few months, how much have you heard or read about generative AI?”, administered to half of the sample in the salience-enhancement condition before any subsequent interventions. Survey weights applied, percentages rounded to the nearest whole number.}
\end{figure}

We now turn to the experimental results. The full regression tables are provided in Appendix \ref{sec:appendixC}. Below, Figure \ref{fig:labeling} presents the effects of AI labeling represented by the difference in group means. H1 predicted that AI labeling would reduce perceived accuracy and this is supported: the perceived accuracy of AI-labeled content is significantly lower than for the unlabeled group (\(\beta = -0.163\) on a 5-point scale, equivalent to 4.1\% of the scale range, \(p < .001\)), consistent with existing studies of AI labeling \parencite{altay_gilardi_2024,longoni_fradkin_cian_pennycook_2022}.  

\begin{figure}[H]
  \centering
  \includegraphics[width=\textwidth]{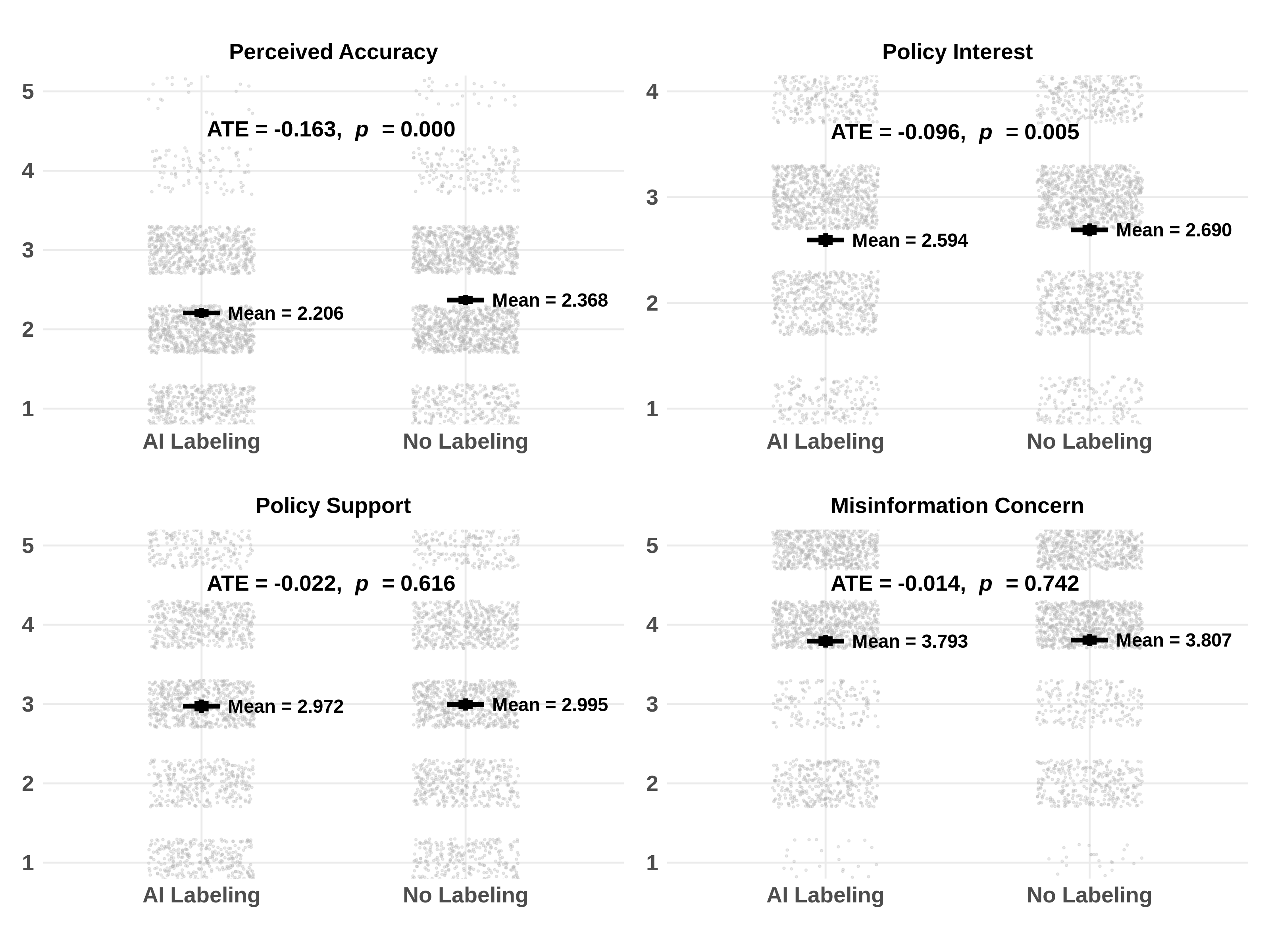}

  \captionsetup{justification=centering}
  \caption{Effects of AI labeling.}

  \captionsetup{justification=raggedright,singlelinecheck=false}
  \caption*{\footnotesize{\textbf{Note}}: Horizontal lines denote group means, and vertical lines denote 95\% (thick) and 99\% (thin) confidence intervals. Light grey points visualize individual responses (jittered vertically and horizontally for visibility). ATE refers to the Average Treatment Effect, estimated as the coefficient on the treatment indicator in a bivariate regression. Two-sided p-values are reported. Survey weights applied.}

  \label{fig:labeling}
\end{figure}

H2 posited that AI labeling would reduce both policy interest and policy support. Consistent with the first part of H2, Figure \ref{fig:labeling} shows that policy interest is lower with AI labeling compared to the no-labeling condition, a statistically significant difference (\(\beta = -0.096\) on a 4-point scale, equivalent to 3.2\% of the scale range, \(p < 0.01\)). In contrast, policy support is not affected (\(\beta = -0.022\) on a 5-point scale, \(p = 0.616\)), providing no support for the second part of H2. One possible explanation for this pattern is that policy interest, as an immediate response to the presented information, may be more susceptible to source-credibility cues such as AI labeling \parencite{petty_cacioppo_1986,winter_bruckner_kraemer_2015}. In contrast, policy support likely reflects more stable ideological predispositions and thus may be less responsive to a single exposure \parencite{lau_heldman_2009,zaller_1992}. H3 proposed that concern about misinformation would be higher in the AI-labeling condition, but this is not supported either (\(\beta = -0.014\) on a 5-point scale, \(p = 0.742\)).

\begin{figure}[H]
  \centering
  \includegraphics[width=\textwidth]{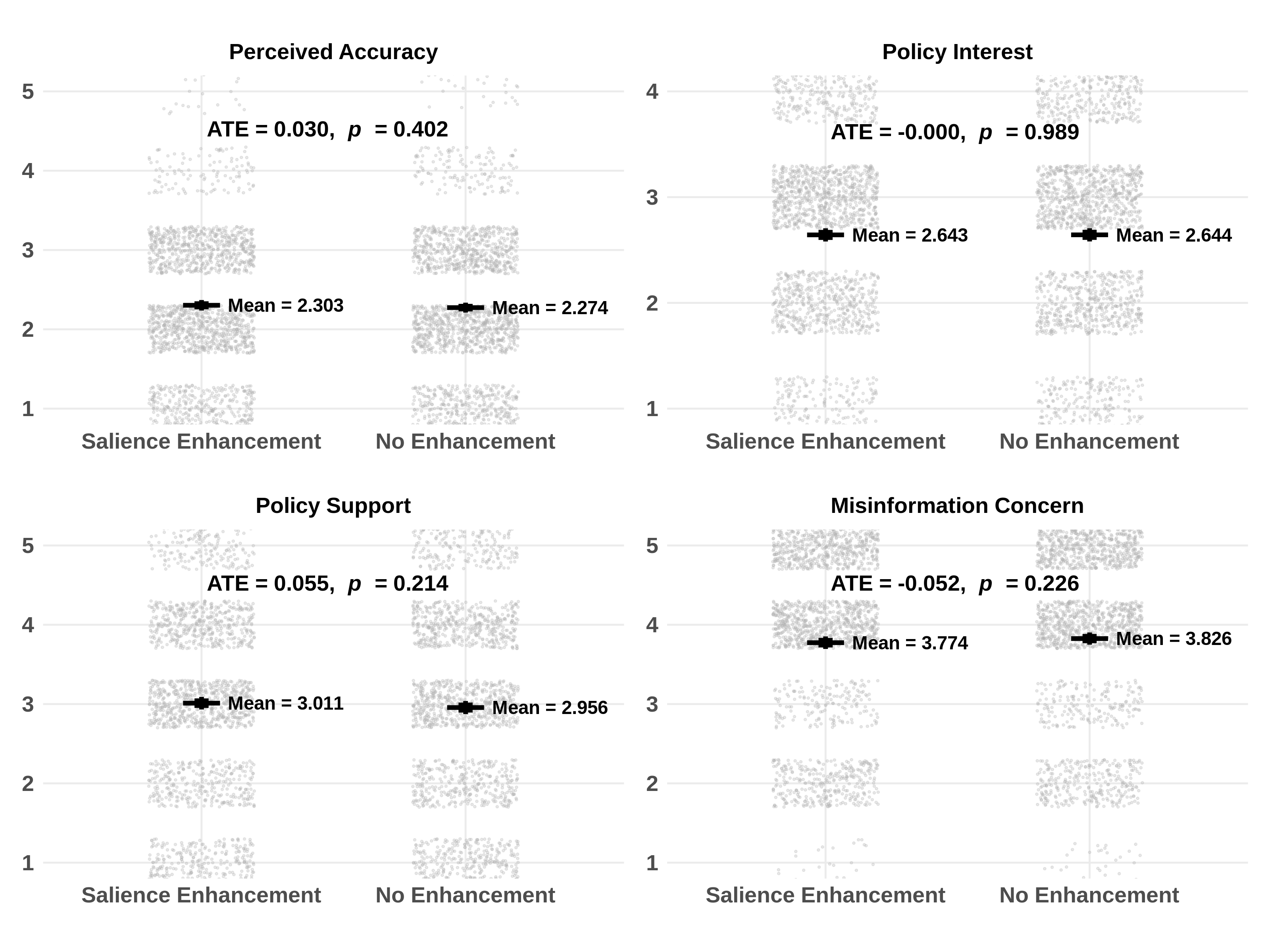}

  \captionsetup{justification=centering}
  \caption{Effects of Salience Enhancement.}

  \captionsetup{justification=raggedright,singlelinecheck=false}
  \caption*{\footnotesize{\textbf{Note}}: Horizontal lines denote group means, and vertical lines denote 95\% (thick) and 99\% (thin) confidence intervals. Light grey points visualize individual responses (jittered vertically and horizontally for visibility). ATE refers to the Average Treatment Effect, estimated as the coefficient on the treatment indicator in a bivariate regression. Two-sided p-values are reported. Survey weights applied.}

  \label{fig:salience}
\end{figure}

H4 posited that perceived accuracy, policy interest, and policy support would be lower, and general misinformation concern higher, when the salience of generative AI is increased. However, none of the group-mean differences for H4 are statistically significant on the four outcomes (Figure \ref{fig:salience}): perceived accuracy (\(\beta = 0.030, p = 0.402\)), policy interest (\(\beta = -0.000, p = 0.989\)), policy support (\(\beta = 0.055, p = 0.214\)), and general misinformation concern (\(\beta = -0.052, p = 0.226\)). This suggests that heightened AI salience is not sufficient on its own to influence public perceptions of online content, at least in the context of public-policy information.

\begin{figure}[H]
  \centering
  \includegraphics[width=\textwidth]{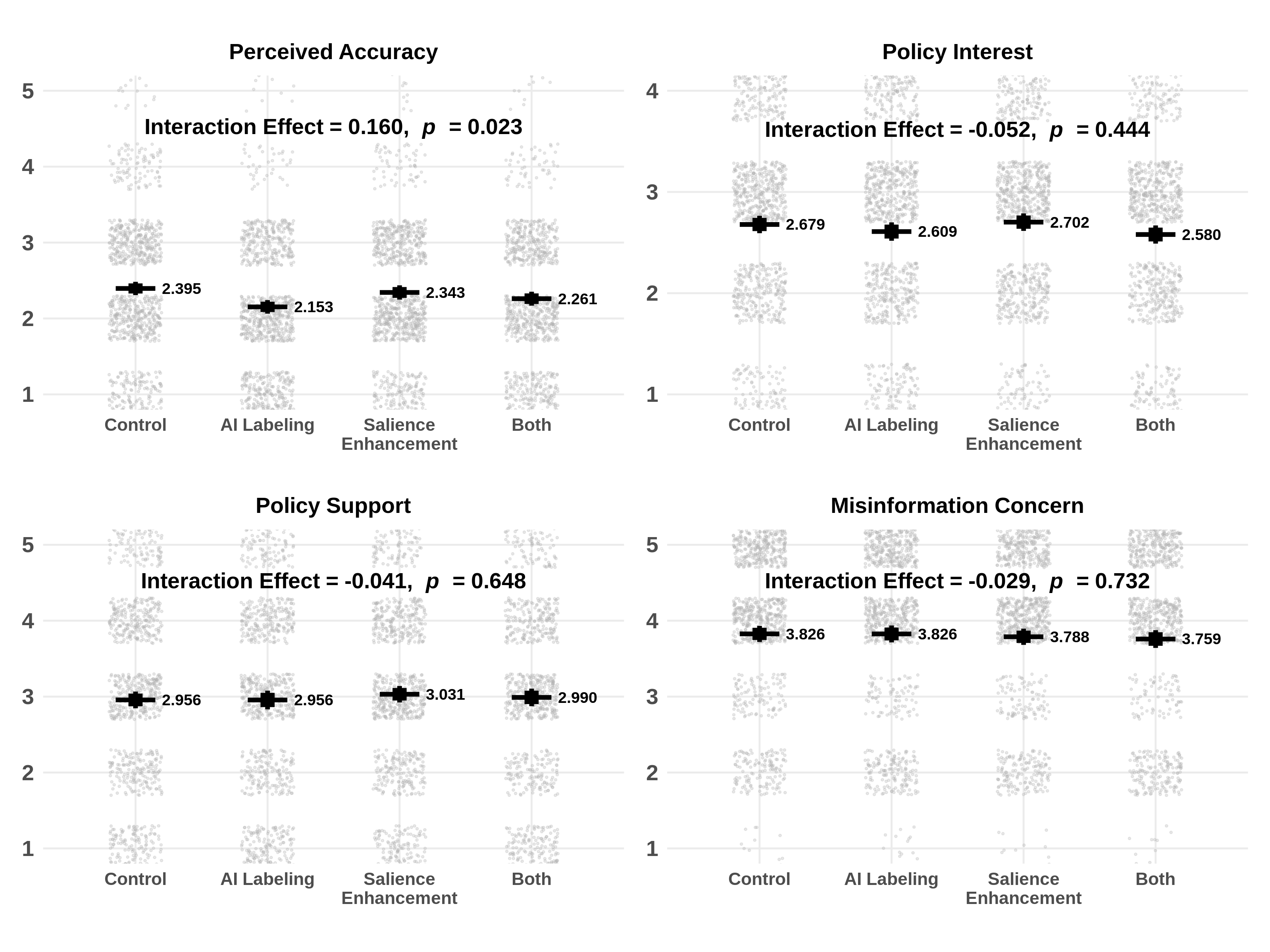}

  \captionsetup{justification=centering}
  \caption{AI Labeling × Salience Enhancement Interaction Effects.}

  \captionsetup{justification=raggedright,singlelinecheck=false}
  \caption*{\footnotesize{\textbf{Note}}: Horizontal lines denote group means, and vertical lines denote 95\% (thick) and 99\% (thin) confidence intervals. Light grey points visualize individual responses (jittered vertically and horizontally for visibility). “Interaction Effect” refers to the coefficient on the AI Labeling × Salience Enhancement interaction term, representing the additional effect of receiving both treatments versus either one alone. Two-sided p-values are reported. Survey weights applied.}

  \label{fig:interaction}
\end{figure}

H5 stated that increasing the salience of generative AI will amplify the negative effect of AI labeling on perceived accuracy, policy interest, and policy support, and will strengthen its effect on heightening general misinformation concern. Figure \ref{fig:interaction} presents these analyses. We find no significant interactions for policy interest (\(\beta = -0.052, p = 0.444\)), policy support (\(\beta = -0.041, p = 0.648\)), or general misinformation concern (\(\beta = -0.029, p = 0.732\)). There is a positive, statistically significant moderating effect of salience enhancement on AI labeling for perceived accuracy (\(\beta = 0.160\) on a 5-point scale, equivalent to 4\% of the scale range, \(p < 0.05\)). Contrary to our prior expectation, however, salience enhancement attenuates the negative effect of AI labeling, increasing perceived accuracy. One possible explanation here is that a neutral salience enhancement served to normalise generative AI in the respondents’ minds, reducing the novelty shock and trust penalty for the subsequent AI labeling. In any event, H5 is not supported.

H6 proposed that perceived accuracy would be lower, while general misinformation concern would be higher, when AI labeling is combined with a one-sided (versus two-sided) framing. Our results show that two-way interaction effects are not statistically significant for perceived accuracy (\(\beta = 0.023, p = 0.739\)) and general misinformation concern (\(\beta = 0.049, p = 0.569\)), failing to support H6. Although not specified in our pre-registered hypotheses, it is worth noting that AI labeling reduced the negative effect of one-sided framing on policy interest (\(\beta = 0.140, p = 0.041\)), implying that labeling may help people evaluate a one-sided narrative more objectively.

Finally, H7 posited that perceived accuracy would be lowest, while general misinformation concern would be highest when the salience of AI is raised, AI labeling is present, and there is a one-sided framing. The three-way interaction was not statistically significant for any of the four outcomes; to conserve space, we do not show the corresponding box-plots.


\section{Discussion}
\label{sec:discussion}

The increasing prevalence of AI-generated content (AIGC) in online environments has led to concerns about a lack of transparency over what content is and is not produced by AI. This ambiguity risks eroding public trust in the quality and credibility of online information. AI labeling has been proposed as a solution to this emerging problem by policymakers in the European Union, the United States, and China \parencite{CAC2025, european_parliament_2024,us_congress_2023}. Yet, AI labeling could have unintended negative consequences; citizens may come to distrust accurate AI-labeled information or, conversely, place an undue level of trust in inaccurate content if it is not labeled as AIGC. Understanding the socio-cognitive impacts of AI labeling is thus a pressing imperative. 

While prior research has primarily focused on the perceived accuracy of AI-labeled content, less is known about whether AI labeling has spillover effects. To advance what we know about the effects of labeling online content as AI-generated, we considered a broader scope of evaluation outcomes using a high-quality, nationally representative probability sample. 

Our findings again show that AI labeling reduces the perceived accuracy of the labeled content. This suggests that some of the mixed results in prior studies may have arisen due to the use of low-quality, opt-in samples, which target sample-specific average treatment effects rather than population average effects. The robustness of this finding suggests that, if implemented as a regulatory requirement, AI labeling may unintentionally serve to undermine public trust in factually accurate information. Policymakers and public communicators should therefore weigh transparency compliance against the potential risk of undermining the perceived credibility of AI-labeled content. From the perspective of algorithm aversion, this pattern is consistent with the idea that explicitly labeling content as AI-generated shifts attention away from the substance of the message toward concerns about the reliability of algorithmic authorship.
By making AI involvement salient in a task that requires evaluative judgment, labeling can lower perceived accuracy even when the underlying information is factually correct.

Beyond perceived accuracy, the spillover effects of AI labeling appear limited. While there is a reduction in expressed policy interest, labeling does not significantly alter policy support or general concerns about online misinformation. The reduced policy interest aligns with evidence that labeling reduces willingness to share labeled online content \parencite{altay_gilardi_2024}. Furthermore, we found no effect of AI labeling on policy support, at least in this UBI policy context. While prior work shows that policy support responds to substantive information about policy effectiveness \parencite{Reynolds2020}, we find that a source-level cue such as AI labeling, absent changes in policy content, does not shift policy preferences. This is encouraging for policymakers, as it suggests labeling may not compromise the effectiveness of policy messaging. With no impact on general misinformation concern, the effects of AI labeling seem more limited than misinformation labeling, which can increase generalized distrust \parencite{hameleers_2023}. The null effect of AI labeling on generalized distrust alleviates a key concern for both policymakers and digital platforms considering the adoption of AI labeling regimes. 

We find no evidence that AI labeling effects are moderated by salience enhancement or one-sided framing. The absence of moderation by message framing suggests that AI labeling effects are not strongly contingent on communicative intent cues, but are instead tied to perceptions of authorship and the production process. The only exception is that salience enhancement does reduce the negative impact of AI labeling on perceived accuracy. This pattern suggests that the initial accuracy penalty associated with AI labeling is partly driven by limited public familiarity with generative AI at baseline, rather than by strong beliefs that AI-generated content is inherently inaccurate.

Taken together, our findings indicate that the scope of AI labeling effects is limited: it reduces perceived accuracy and engagement, while leaving more stable attitudes such as policy support and generalized misinformation concern largely unchanged. This pattern suggests that AI labeling primarily affects evaluative judgments, rather than producing broader downstream attitudinal shifts. Thus, our findings suggest that AI labeling may be implemented as a transparency-promoting measure without significantly undermining the persuasive impact of policy communication or public trust in the wider information environment. 

Although our theoretical discussion centers on algorithm aversion, other mechanisms may plausibly produce similar effects. For example, AI labeling may operate as a less trustworthy source cue \parencite{sundar_2008}, or induce evaluative aversion under uncertainty when public familiarity with generative AI is limited \parencite{zimmermann2024adoption}. Because this preregistered experiment was designed to estimate population-level effects rather than to adjudicate among mediating processes, we do not aim to distinguish between these mechanisms here. Future research could adopt more mechanism-focused designs to disentangle algorithm aversion from related heuristic or uncertainty-based explanations.

We note several limitations of the approach we have taken in this study. First, we focus on a single policy issue (UBI), which may condition how AI labeling affects public evaluations. Future studies could usefully examine AI labeling effects across a broader range of topics and contexts. Second, the experiment relies on a static, one-shot presentation of AI-generated information and does not isolate the effect of AI labeling on text versus images separately. More dynamic or repeated forms of AI labeling that apply to both text and visuals may shape public perceptions differently \parencite{costello_pennycook_rand_2024}. Third, the effects of AI labeling may change as public familiarity with generative AI increases. Thus, understanding how shifts in public awareness shape responses to AI-labeled content is an important direction for future research. A further limitation is that we did not include a conventional standalone manipulation check of perceived salience. Finally, our study relies on a system-specific label (“ChatGPT”) rather than a generic label (“AI-generated”). While this choice reflected the limited public knowledge about generative AI at the time of data collection, system-specific attribution may partly capture attitudes toward the named platform, a possibility that future work could assess by directly comparing generic and system-specific disclosure formats.

Nevertheless, our results suggest that AI labeling effects are currently rather limited in scope. While AI labeling robustly reduces the perceived accuracy of online content and interest in policies described, it has little in the way of a broader effect on more stable attitudes, such as policy support and general misinformation concern. 


\newpage

\printbibliography


\newpage

\appendix
\section{Survey Experimental Materials}

\label{sec:appendixA}

Overall, the flow for the survey experiment is: 1) Salience Enhancement (included or not, based on randomization), 2) a randomized one-sided or two-sided policy article with or without 3) randomized AI Labeling, and 4) final survey evaluation. The following materials reproduce the full experimental content. Grey text shown in the blanket \textcolor{gray}{[gray text]} indicates explanatory notes that were \textbf{not} visible to participants.

\vspace{24pt}

\subsection{Salience Enhancement (Randomized)}
\label{sec:A_salience}
\textcolor{gray}{[Respondents randomized to a salience enhancement condition received the following prompt and question (Q1) at the start of the experiment. All subjects, regardless of salience enhancement, received the last chunk of text introducing the article they would read.]}

\vspace{24pt}

Recently, you may have heard of generative AI. This is a new technology that can quickly produce human-like writing and realistic images. ChatGPT is the most well-known example of generative AI. 

Online news media have started using generative AI to help produce news and lifestyle reports. Many people believe that generative AI will completely change journalism as its use becomes more widespread across the industry. 

\vspace{12pt}

\noindent\textbf{Q1}: Over the past few months, how much have you heard or read about generative AI?

1.	A great deal

2.	Quite a lot

3.	A small amount

4.	Not very much

5.	Nothing at all

6.	Don’t know or not sure

\vspace{12pt}

On the next page, you will be presented with a news article about a policy called Universal Basic Income (UBI). Please read the article carefully before answering some questions about what you have read. 

\vspace{12pt}

\newpage
\subsection{One-Sided Policy Article (Randomized)}
\label{sec:A_one-sided}

\textcolor{gray}{[The next page reproduces the policy article in the one-sided framing condition. The figure caption shown below the image is provided here for documentation only and was \textbf{not} visible to participants.]}

\newpage

\vspace{12pt}

UK · Policy · Posted on 27/05/2024  

\vspace{12pt}

{\large\textbf{UBI: A Brighter Future for All in the UK}}

\begin{figure}[H]
  \includegraphics[width=0.05\textwidth]{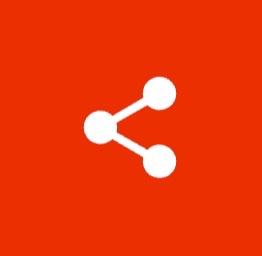}
\end{figure}

\begin{figure}[htbp]
  \includegraphics[width=0.6\textwidth]{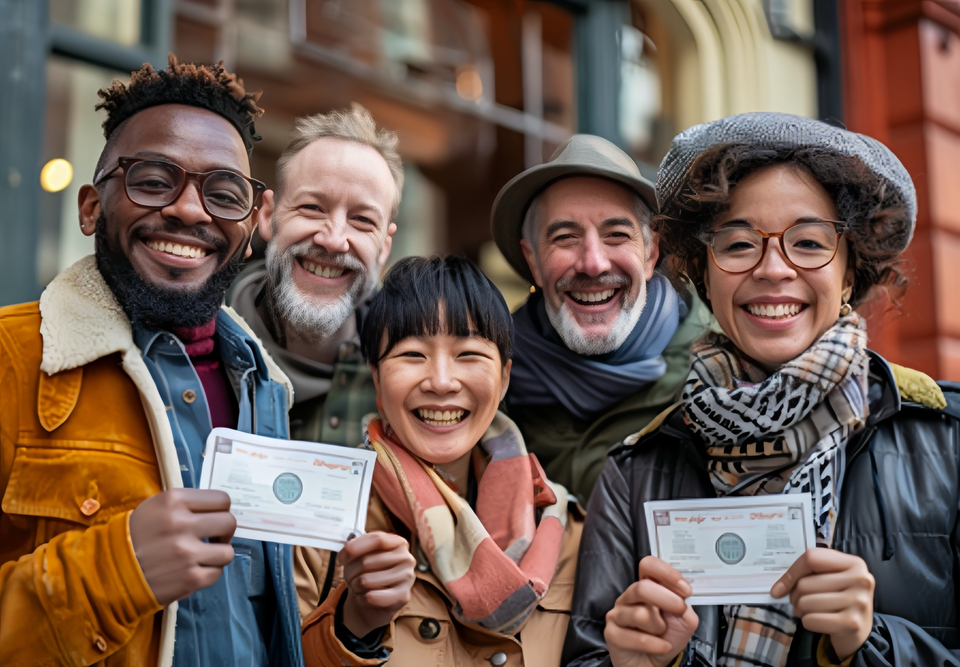}
  \caption{Embedded image for the policy article with one-sided framing.} 
  \label{fig:onesided}
\end{figure}

\textbf{A pioneering UBI pilot in Wales sparks hope for a financially inclusive UK.}

Imagine a future where financial worries are a thing of the past, where every person has enough money to cover their basic needs each month. This isn't just a dream—it's the promise of Universal Basic Income (UBI).

The Welsh Government is leading the charge with an exciting UBI trial, where a group of 500 people are now receiving £1,600 before tax every month. They receive this amount, whatever their working status, and are free to use it however they want.

Following successful trials in the USA, Finland, and Spain, it seems that UBI may offer a path to a future without poverty. Choices between essentials like heating or eating will disappear, giving everyone the necessary foundation to thrive.

Advocates of UBI argue that local economies will also thrive as people have more to spend, driving growth and creating jobs in communities across the UK. UBI may even have the potential to tackle poverty while streamlining the welfare system.

The promise of UBI is compelling: it represents not just a financial policy but a step towards a society that values prosperity for all in the UK.

\newpage

\subsection{Two-Sided Policy Article (Randomized)}
\label{sec:A_two-sided}

\textcolor{gray}{[The next page reproduces the policy article in the one-sided framing condition. The figure caption shown below the image is provided here for documentation only and was \textbf{not} visible to participants. Each respondent read either this version of the article or the version in Section \ref{sec:A_one-sided}.]}

\newpage

\vspace{12pt}

UK · Policy · Posted on 27/05/2024    

\vspace{12pt}

{\large\textbf{UBI: A New Horizon or a Step Too Far?}}

\begin{figure}[H]
  \includegraphics[width=0.05\textwidth]{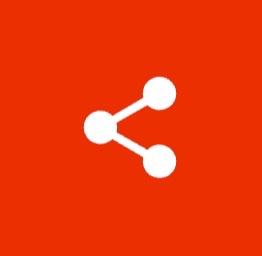}
\end{figure}
 
\begin{figure}[H]
  \includegraphics[width=0.6\textwidth]{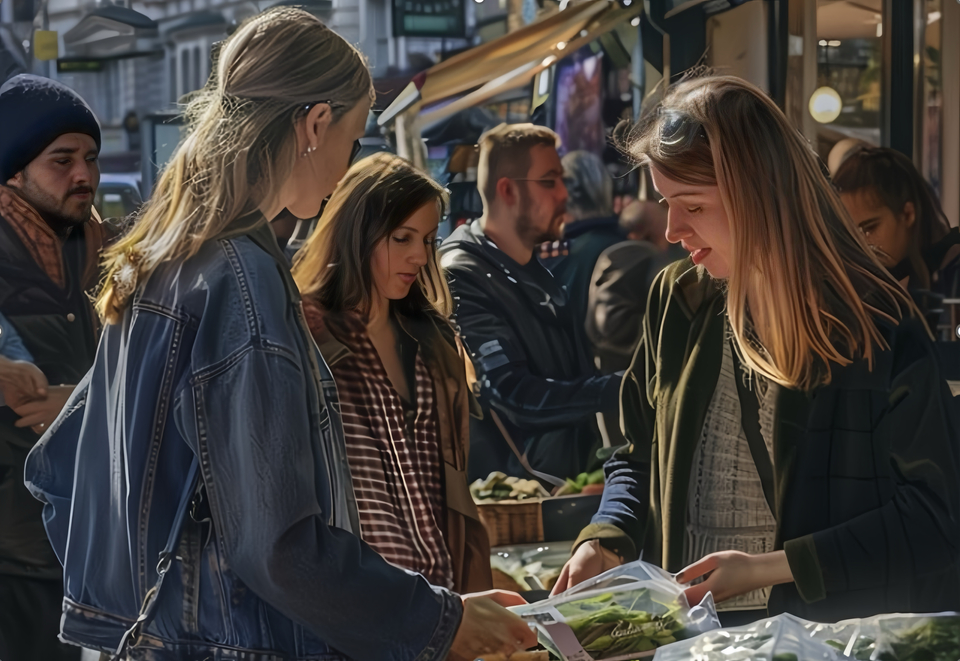}
    \caption{Embedded image for the policy article with two-sided framing.} 
\end{figure}

\textbf{A new pilot project in Wales sets the stage for debate about universal basic income in the UK.}

What if the government provided everyone in the UK with enough money to live on each month, whether they work or not? This is the aim of universal basic income (UBI). 

The Welsh Government has recently launched a UBI trial. The plan gives £1,600 a month before tax to 500 people to provide them with an income close to the national living wage. They receive this amount, whatever their working status, and are free to use it however they want.

The pilot’s participants are representative of the local population, including individuals with disabilities. This UBI initiative, informed by extensive community consultations, not only looks at UBI's economic impact but also it is potential to enhance mental and physical well-being.

Advocates of UBI say it will help to tackle poverty by ensuring a safety net for all while supporting job creation. It may also stimulate demand in local areas and drive inclusive economic growth. 

Critics, on the other hand, are concerned that UBI will be extremely expensive, discourage people from finding jobs, and may worsen the UK economy by pushing up the cost of living further. 

\newpage

\subsection{AI Labeling (Randomized)} 
\label{sec:A_labeling}

\textcolor{gray}{[The following text and logo were included at the bottom of the articles when respondents were in the AI labeling condition.]}

\vspace{12pt}

\noindent\textbf{\textit{Disclaimer: This report was generated by ChatGPT, an artificial intelligence.}}

\begin{figure}[H]
  \includegraphics[width=0.08\textwidth]{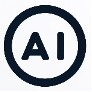}
\end{figure}

\vspace{12pt}

\subsection{Survey Evaluation}
\label{sec:A_survey}

\textcolor{gray}{[All participants answer the same set of survey questions below.]}

\vspace{12pt}

\noindent\textbf{Q2}: How much confidence do you have in the accuracy of the information presented in the article? 

1.	No confidence

2.	Little confidence

3.	Some confidence

4.	Confidence

5.	High confidence

\noindent\textbf{Q3}: How interested would you be in learning more about Universal Basic Income?

1.	Very interested

2.	Quite interested

3.	Not very interested

4.	Not at all interested

\noindent\textbf{Q4}: If there were a proposal to introduce UBI in the UK, how likely would you be to support it? 

1.	Very unlikely

2.	Unlikely

3.	Neither likely nor unlikely

4.	Likely 

5.	Very likely

\noindent\textbf{Q5}: Generally speaking, how concerned are you about misinformation in online news media?

1.	Not at all concerned

2.	Slightly concerned

3.	Neither concerned nor unconcerned

4.	Quite concerned

5.	Very concerned

\noindent\textcolor{gray}{[Note: No “Don’t know” or “Prefer not to answer” options were provided. Participants were required to select one of the response categories above.]}


\newpage

\section{Regression Specifications}

\label{sec:appendixB}

The main results reported in the paper are estimated with weighted least squares (WLS) using survey weights created by the survey provider Verian. We estimate a range of specifications: 

\begin{equation}
\label{eq:model1}
\begin{aligned}
Y_{\text{outcome}}  \;=\;&
  \beta_{0}
+ \beta_{1}\,\text{(Labeling)}
+ \varepsilon
\end{aligned}
\end{equation}

\begin{center}
Specification 1: Effects of AI Labeling.
\end{center}

\begin{equation}
\label{eq:model2}
\begin{aligned}
Y_{\text{outcome}}  \;=\;&
  \beta_{0}
+ \beta_{1}\,\text{(Salience)}
+ \varepsilon
\end{aligned}
\end{equation}

\begin{center}
Specification 2: Effects of Salience Enhancement.
\end{center}

\begin{equation}
\label{eq:model3}
\begin{aligned}
Y_{\text{outcome}}  \;=\;&
  \beta_{0}
+ \beta_{1}\,\text{(One-side)}
+ \varepsilon
\end{aligned}
\end{equation}

\begin{center}
Specification 3: Effects of One-sided Framing.
\end{center}

\begin{equation}
\label{eq:model4}
\begin{aligned}
Y_{\text{outcome}}  \;=\;&
  \beta_{0}
+ \beta_{1}\,\text{(Labeling)}
+ \beta_{2}\,\text{(Salience)} \\
&+ \beta_{3}\,\text{(Labeling}\!\times\!\text{Salience)}
+ \varepsilon
\end{aligned}
\end{equation}

\begin{center}
Specification 4: Effects of AI Labeling $\times$ Salience Enhancement.
\end{center}

\begin{equation}
\label{eq:model5}
\begin{aligned}
Y_{\text{outcome}}  \;=\;&
  \beta_{0}
+ \beta_{1}\,\text{(Labeling)}
+ \beta_{3}\,\text{(One-side)}  \\
&+ \beta_{6}\,\text{(Labeling}\!\times\!\text{One-side)}
+ \varepsilon
\end{aligned}
\end{equation}

\begin{center}
Specification 5: Effects of AI Labeling $\times$ One-sided Framing.
\end{center}

\begin{equation}
\label{eq:model6}
\begin{aligned}
Y_{\text{outcome}}  \;=\;&
  \beta_{0}
+ \beta_{2}\,\text{(Salience)}
+ \beta_{3}\,\text{(One-side)}  \\
&+ \beta_{5}\,\text{(Salience}\!\times\!\text{One-side)}   
+ \varepsilon
\end{aligned}
\end{equation}

\begin{center}
Specification 6: Effects of Salience Enhancement $\times$ One-sided Framing.
\end{center}

\begin{equation}
\label{eq:model7}
\begin{aligned}
Y_{\text{outcome}}  \;=\;&
  \beta_{0}
+ \beta_{1}\,\text{(Labeling)}
+ \beta_{2}\,\text{(Salience)}
+ \beta_{3}\,\text{(One-side)}  \\
&+ \beta_{4}\,\text{(Labeling}\!\times\!\text{Salience)}
+ \beta_{5}\,\text{(Salience}\!\times\!\text{One-side)}    
+ \beta_{6}\,\text{(Labeling}\!\times\!\text{One-side)} \\
&+ \beta_{7}\,\text{(Labeling}\!\times\!\text{Salience}\!\times\!\text{One-side)}
+ \varepsilon
\end{aligned}
\end{equation}

\begin{center}
Specification 7: Effects of AI Labeling $\times$ Salience Enhancement $\times$ One-sided Framing.
\end{center}

Where $Y_{\text{outcome}}$ denotes any of the four dependent variables
(Perceived Accuracy, Policy Interest, Policy Support, General Misinformation Concern),
$\beta_0$ is an intercept, and $\varepsilon$ is a stochastic error term, and the remaining $\beta$ coefficients are main or interaction effects of the three experimentally manipulated factors: AI~Labeling, Salience Enhancement, and One-sided Framing.

In general: 
\begin{enumerate}
    \item Specification 1 is used to test H1–H3, Specification 2 to test H4, Specification 4  to test H5, Specification 5  to test H6, and Specification 7  to test H7.
    \item Specifications 1–3, 4–6, and 7 correspond respectively to Tables \ref{tab:C1}, \ref{tab:D1}, and \ref{tab:E1}, \ref{tab:C2}, \ref{tab:D2}, and \ref{tab:E2}, as well as \ref{tab:C3}, \ref{tab:D3}, and \ref{tab:E3}, respectively.
\end{enumerate}


\newpage

\section{OLS Weighted Regression Results}

\label{sec:appendixC}
\setcounter{table}{0}
\renewcommand{\thetable}{B\arabic{table}}

See Appendix B for the mapping of hypotheses to specifications and tables.


\setcounter{table}{0}
\renewcommand{\thetable}{C\arabic{table}}
\begin{table}[htbp]
\centering
\caption{Three models with individual effects of each treatment}
\label{tab:C1}
\resizebox{\textwidth}{!}{%
\newcolumntype{C}{>{\centering\arraybackslash}S[table-format=-1.3,table-number-alignment=center]}
\begin{tabular}{l CCCC}
\toprule
\textbf{Conditions} & {\textbf{Perceived Accuracy}} & {\textbf{UBI Interest}} & {\textbf{UBI Support}} & {\textbf{Misinformation Concern}} \\
\midrule
\multicolumn{5}{l}{\textit{AI Labeling only}} \\
AI Labeling & -0.163\textsuperscript{***} & -0.096\textsuperscript{**} & -0.022 & -0.014 \\
 & {(0.035)} & {(0.034)} & {(0.045)} & {(0.043)} \\
\hdashline
\multicolumn{5}{l}{\textit{Salience Enhancement only}} \\
Salience Enhancement & 0.030 & -0.000 & 0.055 & -0.052 \\
 & {(0.035)} & {(0.034)} & {(0.044)} & {(0.043)} \\
\hdashline
\multicolumn{5}{l}{\textit{One-sided Framing only}} \\
One-sided Framing & -0.101\textsuperscript{**} & 0.114\textsuperscript{***} & 0.306\textsuperscript{***} & 0.066 \\
 & {(0.035)} & {(0.034)} & {(0.044)} & {(0.043)} \\
\bottomrule
\end{tabular}%
}
\vspace{0.5ex}
\begin{minipage}{\textwidth}
\footnotesize
Note: Each model (with survey weights) includes a single treatment variable only. The format is Coefficient (Robust Standard Error). Significance: $^{***}p < 0.001$, $^{**}p < 0.01$, $^{*}p < 0.05$. In subsequent tables, AI Labeling, Salience Enhancement, and One-sided Framing are abbreviated as Labeling, Salience, and One-sided.
\end{minipage}
\end{table}


\begin{table}[htbp]
\centering
\caption{Three models with effects of any two treatments and their interaction}
\label{tab:C2}
\resizebox{\textwidth}{!}{%
\newcolumntype{C}{>{\centering\arraybackslash}S[table-format=-1.3,table-number-alignment=center]}
\begin{tabular}{l CCCC}
\toprule
\textbf{Conditions} & {\textbf{Perceived Accuracy}} & {\textbf{UBI Interest}} & {\textbf{UBI Support}} & {\textbf{Misinformation Concern}} \\
\midrule
\multicolumn{5}{l}{\textit{Labeling and Salience}} \\
Labeling & -0.243\textsuperscript{***} & -0.070 & -0.001 & -0.001 \\
 & {(0.048)} & {(0.048)} & {(0.064)} & {(0.059)} \\
Salience & -0.053 & 0.023 & 0.075 & -0.038 \\
 & {(0.050)} & {(0.047)} & {(0.060)} & {(0.058)} \\
Labeling $\times$ Salience & 0.160\textsuperscript{*} & -0.052 & -0.041 & -0.029 \\
 & {(0.070)} & {(0.068)} & {(0.089)} & {(0.086)} \\
\hdashline
\multicolumn{5}{l}{\textit{Labeling and One-sided}} \\
Labeling & -0.178\textsuperscript{***} & -0.164\textsuperscript{***} & -0.021 & -0.037 \\
 & {(0.049)} & {(0.049)} & {(0.065)} & {(0.061)} \\
One-sided & -0.117\textsuperscript{*} & 0.042 & 0.298\textsuperscript{***} & 0.042 \\
 & {(0.050)} & {(0.047)} & {(0.060)} & {(0.059)} \\
Labeling $\times$ One-sided & 0.023 & 0.140\textsuperscript{*} & 0.015 & 0.049 \\
 & {(0.070)} & {(0.068)} & {(0.089)} & {(0.086)} \\
\hdashline
\multicolumn{5}{l}{\textit{Salience and One-sided}} \\
Salience & 0.040 & 0.023 & 0.096 & -0.090 \\
 & {(0.049)} & {(0.049)} & {(0.065)} & {(0.061)} \\
One-sided & -0.092 & 0.138\textsuperscript{**} & 0.348\textsuperscript{***} & 0.030 \\
 & {(0.049)} & {(0.048)} & {(0.064)} & {(0.059)} \\
Salience $\times$ One-sided & -0.018 & -0.048 & -0.085 & 0.074 \\
 & {(0.071)} & {(0.069)} & {(0.089)} & {(0.085)} \\
\bottomrule
\end{tabular}%
}
\vspace{0.5ex}
\begin{minipage}{\textwidth}
\footnotesize
Note: Each model (with survey weights) includes two of the three treatments with their interaction term. The format is Coefficient (Robust Standard Error). Significance: $^{***}p < 0.001$, $^{**}p < 0.01$, $^{*}p < 0.05$.
\end{minipage}
\end{table}


\clearpage
\vspace*{0cm}
\begin{table}[H] 
\centering
\caption{The full model with main, two-way, and three-way interaction effects}
\label{tab:C3}
\resizebox{\textwidth}{!}{%
\newcolumntype{C}{>{\centering\arraybackslash}S[table-format=-1.3,table-number-alignment=center]}
\begin{tabular}{l CCCC}
\toprule
\textbf{Conditions} & {\textbf{Perceived Accuracy}} & {\textbf{UBI Interest}} & {\textbf{UBI Support}} & {\textbf{Misinformation Concern}} \\
\midrule
Labeling & -0.238\textsuperscript{***} & -0.124 & -0.008 & -0.057 \\
 & {(0.069)} & {(0.069)} & {(0.091)} & {(0.081)} \\
Salience & -0.020 & 0.066 & 0.111 & -0.111 \\
 & {(0.069)} & {(0.067)} & {(0.090)} & {(0.087)} \\
One-sided & -0.088 & 0.084 & 0.341\textsuperscript{***} & -0.026 \\
 & {(0.068)} & {(0.067)} & {(0.085)} & {(0.082)} \\
Labeling $\times$ Salience & 0.121 & -0.080 & -0.028 & 0.042 \\
 & {(0.098)} & {(0.097)} & {(0.129)} & {(0.122)} \\
Labeling $\times$ One-sided & -0.008 & 0.107 & 0.013 & 0.112 \\
 & {(0.097)} & {(0.097)} & {(0.127)} & {(0.119)} \\
Salience $\times$ One-sided & -0.055 & -0.084 & -0.089 & 0.137 \\
 & {(0.099)} & {(0.095)} & {(0.120)} & {(0.117)} \\
Labeling $\times$ Salience $\times$ One-sided & 0.064 & 0.064 & 0.008 & -0.129 \\
 & {(0.140)} & {(0.137)} & {(0.177)} & {(0.172)} \\
\bottomrule
\end{tabular}%
}
\vspace{0.5ex}
\begin{minipage}{\textwidth}
\footnotesize
Note: The model (with survey weights) includes all three treatments, all three two-way interactions, and the three-way interaction. The format is Coefficient (Robust Standard Error). Significance: $^{***}p < 0.001$, $^{**}p < 0.01$, $^{*}p < 0.05$.
\end{minipage}
\end{table}

\clearpage

\newpage

\section{OLS Unweighted Regression Results}

\label{sec:appendixD}
\setcounter{table}{0}
\renewcommand{\thetable}{B\arabic{table}}

See Appendix B for the mapping of hypotheses to specifications and tables.


\setcounter{table}{0}
\renewcommand{\thetable}{D\arabic{table}}
\begin{table}[htbp]
\centering
\caption{Three models with individual effects of each treatment}
\label{tab:D1}
\resizebox{\textwidth}{!}{%
\newcolumntype{C}{>{\centering\arraybackslash}S[table-format=-1.3,table-number-alignment=center]}
\begin{tabular}{l CCCC}
\toprule
\textbf{Conditions} & {\textbf{Perceived Accuracy}} & {\textbf{UBI Interest}} & {\textbf{UBI Support}} & {\textbf{Misinformation Concern}} \\
\midrule
\multicolumn{5}{l}{\textit{AI Labeling only}} \\
AI Labeling & -0.197\textsuperscript{***} & -0.092\textsuperscript{**} & -0.038 & -0.005 \\
 & {(0.029)} & {(0.030)} & {(0.040)} & {(0.036)} \\
 \hdashline
\multicolumn{5}{l}{\textit{Salience Enhancement only}} \\
Salience & -0.001 & -0.002 & 0.020 & -0.033 \\
 & {(0.029)} & {(0.030)} & {(0.040)} & {(0.036)} \\
\hdashline
\multicolumn{5}{l}{\textit{One-sided Framing only}} \\
One-sided Framing & -0.143\textsuperscript{***} & 0.134\textsuperscript{***} & 0.336\textsuperscript{***} & 0.087\textsuperscript{*} \\
 & {(0.029)} & {(0.029)} & {(0.039)} & {(0.036)} \\
\bottomrule
\end{tabular}%
}
\vspace{0.5ex}
\begin{minipage}{\textwidth}
\footnotesize
Note: Each model (with survey weights) includes a single treatment variable only. The format is Coefficient (Robust Standard Error). Significance: $^{***}p < 0.001$, $^{**}p < 0.01$, $^{*}p < 0.05$. In subsequent tables, AI Labeling, Salience Enhancement, and One-sided Framing are abbreviated as Labeling, Salience, and One-sided.
\end{minipage}
\end{table}


\begin{table}[htbp]
\centering
\caption{Three models with effects of any two treatments and their interaction}
\label{tab:D2}
\resizebox{\textwidth}{!}{%
\newcolumntype{C}{>{\centering\arraybackslash}S[table-format=-1.3,table-number-alignment=center]}
\begin{tabular}{l CCCC}
\toprule
\textbf{Conditions} & {\textbf{Perceived Accuracy}} & {\textbf{UBI Interest}} & {\textbf{UBI Support}} & {\textbf{Misinformation Concern}} \\
\midrule
\multicolumn{5}{l}{\textit{Labeling and Salience}} \\
Labeling & -0.309\textsuperscript{***} & -0.091\textsuperscript{*} & -0.045 & -0.009 \\
 & {(0.042)} & {(0.042)} & {(0.057)} & {(0.050)} \\
Salience & -0.116\textsuperscript{**} & -0.003 & 0.011 & -0.037 \\
 & {(0.041)} & {(0.041)} & {(0.055)} & {(0.049)} \\
Labeling $\times$ Salience & 0.226\textsuperscript{***} & -0.002 & 0.016 & 0.007 \\
 & {(0.058)} & {(0.059)} & {(0.080)} & {(0.071)} \\
\hdashline
\multicolumn{5}{l}{\textit{Labeling and One-sided}} \\
Labeling & -0.211\textsuperscript{***} & -0.131\textsuperscript{**} & 0.002 & -0.029 \\
 & {(0.042)} & {(0.042)} & {(0.057)} & {(0.051)} \\
One-sided & -0.159\textsuperscript{***} & 0.091\textsuperscript{*} & 0.366\textsuperscript{***} & 0.062 \\
 & {(0.041)} & {(0.041)} & {(0.054)} & {(0.049)} \\
Labeling $\times$ One-sided & 0.021 & 0.083 & -0.064 & 0.051 \\
 & {(0.058)} & {(0.059)} & {(0.079)} & {(0.071)} \\
\hdashline
\multicolumn{5}{l}{\textit{Salience and One-sided}} \\
Salience & 0.009 & 0.011 & 0.042 & -0.049 \\
 & {(0.042)} & {(0.042)} & {(0.057)} & {(0.051)} \\
One-sided & -0.135\textsuperscript{**} & 0.148\textsuperscript{***} & 0.360\textsuperscript{***} & 0.073 \\
 & {(0.042)} & {(0.042)} & {(0.056)} & {(0.050)} \\
Salience $\times$ One-sided & -0.017 & -0.029 & -0.050 & 0.028 \\
 & {(0.058)} & {(0.059)} & {(0.079)} & {(0.071)} \\
\bottomrule
\end{tabular}%
}
\vspace{0.5ex}
\begin{minipage}{\textwidth}
\footnotesize
Note: Each model (with survey weights) includes two of the three treatments with their interaction term. The format is Coefficient (Robust Standard Error). Significance: $^{***}p < 0.001$, $^{**}p < 0.01$, $^{*}p < 0.05$.
\end{minipage}
\end{table}


\clearpage
\vspace*{0cm}
\begin{table}[H] 
\centering
\caption{The full model with main, two-way, and three-way interaction effects}
\label{tab:D3}
\resizebox{\textwidth}{!}{%
\newcolumntype{C}{>{\centering\arraybackslash}S[table-format=-1.3,table-number-alignment=center]}
\begin{tabular}{l CCCC}
\toprule
\textbf{Conditions} & {\textbf{Perceived Accuracy}} & {\textbf{UBI Interest}} & {\textbf{UBI Support}} & {\textbf{Misinformation Concern}} \\
\midrule
Labeling & -0.289\textsuperscript{***} & -0.132\textsuperscript{*} & 0.003 & -0.069 \\
 & {(0.061)} & {(0.061)} & {(0.082)} & {(0.071)} \\
Salience & -0.069 & 0.012 & 0.043 & -0.090 \\
 & {(0.059)} & {(0.059)} & {(0.080)} & {(0.073)} \\
One-sided & -0.114 & 0.109 & 0.410\textsuperscript{***} & 0.014 \\
 & {(0.061)} & {(0.059)} & {(0.078)} & {(0.070)} \\
Labeling $\times$ Salience & 0.159 & 0.002 & -0.002 & 0.083 \\
 & {(0.084)} & {(0.085)} & {(0.115)} & {(0.103)} \\
Labeling $\times$ One-sided & -0.039 & 0.080 & -0.099 & 0.119 \\
 & {(0.083)} & {(0.085)} & {(0.113)} & {(0.100)} \\
Salience $\times$ One-sided & -0.080 & -0.035 & -0.087 & 0.098 \\
 & {(0.082)} & {(0.082)} & {(0.109)} & {(0.099)} \\
Labeling $\times$ Salience $\times$ One-sided & 0.117 & 0.005 & 0.071 & -0.140 \\
 & {(0.116)} & {(0.118)} & {(0.158)} & {(0.143)} \\
\bottomrule
\end{tabular}%
}
\vspace{0.5ex}
\begin{minipage}{\textwidth}
\footnotesize
Note: The model (with survey weights) includes all three treatments, all three two-way interactions, and the three-way interaction. The format is Coefficient (Robust Standard Error). Significance: $^{***}p < 0.001$, $^{**}p < 0.01$, $^{*}p < 0.05$.
\end{minipage}
\end{table}


\clearpage

\newpage

\section{OLR Weighted Regression Results}
\label{sec:appendixE}
\setcounter{table}{0}
\renewcommand{\thetable}{E\arabic{table}}

See Appendix B for the mapping of hypotheses to specifications and tables.


\begin{table}[htbp]
\centering
\caption{Three models with individual effects of each treatment}
\label{tab:E1}
\resizebox{\textwidth}{!}{%
\newcolumntype{C}{>{\centering\arraybackslash}S[table-format=-1.3,table-number-alignment=center]}
\begin{tabular}{l CCCC}
\toprule
\textbf{Conditions} & {\textbf{Perceived Accuracy}} & {\textbf{UBI Interest}} & {\textbf{UBI Support}} & {\textbf{Misinformation Concern}} \\
\midrule
\multicolumn{5}{l}{\textit{AI Labeling only}} \\
AI Labeling & -0.334\textsuperscript{***} & -0.182\textsuperscript{**} & -0.029 & 0.008 \\
 & {(0.059)} & {(0.069)} & {(0.083)} & {(0.103)} \\
\hdashline
\multicolumn{5}{l}{\textit{Salience Enhancement only}} \\
Salience Enhancement & 0.070 & -0.004 & 0.076 & -0.082 \\
 & {(0.064)} & {(0.072)} & {(0.085)} & {(0.104)} \\
\hdashline
\multicolumn{5}{l}{\textit{One-sided Framing only}} \\
One-sided Framing & -0.214\textsuperscript{***} & 0.216\textsuperscript{**} & 0.445\textsuperscript{***} & 0.104 \\
 & {(0.061)} & {(0.071)} & {(0.086)} & {(0.103)} \\
\bottomrule
\end{tabular}%
}
\vspace{0.5ex}
\begin{minipage}{\textwidth}
\footnotesize
Note: Each model (with survey weights) includes a single treatment variable only. The format is Coefficient (Robust Standard Error). Significance: $^{***}p < 0.001$, $^{**}p < 0.01$, $^{*}p < 0.05$. In subsequent tables, AI Labeling, Salience Enhancement, and One-sided Framing are abbreviated as Labeling, Salience, and One-sided.
\end{minipage}
\end{table}


\begin{table}[htbp]
\centering
\caption{Three models with effects of any two treatments and their interaction}
\label{tab:E2}
\resizebox{\textwidth}{!}{%
\newcolumntype{C}{>{\centering\arraybackslash}S[table-format=-1.3,table-number-alignment=center]}
\begin{tabular}{l CCCC}
\toprule
\textbf{Conditions} & {\textbf{Perceived Accuracy}} & {\textbf{UBI Interest}} & {\textbf{UBI Support}} & {\textbf{Misinformation Concern}} \\
\midrule
\multicolumn{5}{l}{\textit{Labeling and Salience}} \\
Labeling & -0.485\textsuperscript{***} & -0.139 & -0.000 & 0.018 \\
 & {(0.080)} & {(0.098)} & {(0.117)} & {(0.146)} \\
Salience & -0.090 & 0.034 & 0.102 & -0.070 \\
 & {(0.093)} & {(0.104)} & {(0.121)} & {(0.148)} \\
Labeling $\times$ Salience & 0.304\textsuperscript{*} & -0.087 & -0.054 & -0.024 \\
 & {(0.151)} & {(0.173)} & {(0.206)} & {(0.253)} \\
\hdashline
\multicolumn{5}{l}{\textit{Labeling and One-sided}} \\
Labeling & -0.361\textsuperscript{***} & -0.309\textsuperscript{***} & -0.019 & -0.068 \\
 & {(0.084)} & {(0.094)} & {(0.111)} & {(0.142)} \\
One-sided & -0.243\textsuperscript{**} & 0.087 & 0.442\textsuperscript{***} & 0.027 \\
 & {(0.089)} & {(0.099)} & {(0.118)} & {(0.143)} \\
Labeling $\times$ One-sided & 0.042 & 0.257 & 0.004 & 0.160 \\
 & {(0.148)} & {(0.171)} & {(0.204)} & {(0.251)} \\
\hdashline
\multicolumn{5}{l}{\textit{Salience and One-sided}} \\
Salience & 0.104 & 0.055 & 0.152 & -0.118 \\
 & {(0.091)} & {(0.101)} & {(0.118)} & {(0.145)} \\
One-sided & -0.181\textsuperscript{*} & 0.275\textsuperscript{**} & 0.520\textsuperscript{***} & 0.074 \\
 & {(0.083)} & {(0.101)} & {(0.120)} & {(0.145)} \\
Salience $\times$ One-sided & -0.064 & -0.116 & -0.149 & 0.066 \\
 & {(0.152)} & {(0.175)} & {(0.208)} & {(0.253)} \\
\bottomrule
\end{tabular}%
}
\vspace{0.5ex}
\begin{minipage}{\textwidth}
\footnotesize
Note: Each model (with survey weights) includes two of the three treatments with their interaction term. The format is Coefficient (Robust Standard Error). Significance: $^{***}p < 0.001$, $^{**}p < 0.01$, $^{*}p < 0.05$.
\end{minipage}
\end{table}


\clearpage
\vspace*{0cm}
\begin{table}[H] 
\centering
\caption{The full model with main, two-way, and three-way interaction effects}
\label{tab:E3}
\resizebox{\textwidth}{!}{%
\newcolumntype{C}{>{\centering\arraybackslash}S[table-format=-1.3,table-number-alignment=center]}
\begin{tabular}{l CCCC}
\toprule
\textbf{Conditions} & {\textbf{Perceived Accuracy}} & {\textbf{UBI Interest}} & {\textbf{UBI Support}} & {\textbf{Misinformation Concern}} \\
\midrule
Labeling & -0.458\textsuperscript{***} & -0.244 & 0.011 & -0.087 \\
 & {(0.113)} & {(0.133)} & {(0.152)} & {(0.202)} \\
Salience & 0.004 & 0.123 & 0.183 & -0.138 \\
 & {(0.134)} & {(0.155)} & {(0.172)} & {(0.212)} \\
One-sided & -0.154 & 0.176 & 0.529\textsuperscript{**} & -0.032 \\
 & {(0.120)} & {(0.140)} & {(0.162)} & {(0.199)} \\
Labeling $\times$ Salience & 0.193 & -0.130 & -0.062 & 0.042 \\
 & {(0.214)} & {(0.243)} & {(0.281)} & {(0.354)} \\
Labeling $\times$ One-sided & -0.054 & 0.201 & -0.017 & 0.210 \\
 & {(0.200)} & {(0.241)} & {(0.284)} & {(0.353)} \\
Salience $\times$ One-sided & -0.168 & -0.175 & -0.173 & 0.124 \\
 & {(0.221)} & {(0.252)} & {(0.291)} & {(0.356)} \\
Labeling $\times$ Salience $\times$ One-sided & 0.192 & 0.106 & 0.045 & -0.111 \\
 & {(0.324)} & {(0.375)} & {(0.443)} & {(0.544)} \\
\bottomrule
\end{tabular}%
}
\vspace{0.5ex}
\begin{minipage}{\textwidth}
\footnotesize
Note: The model (with survey weights) includes all three treatments, all three two-way interactions, and the three-way interaction. The format is Coefficient (Robust Standard Error). Significance: $^{***}p < 0.001$, $^{**}p < 0.01$, $^{*}p < 0.05$.
\end{minipage}
\end{table}

\clearpage

\end{document}